\documentclass[conference]{IEEEtran}
\usepackage{amsmath,amssymb,amsfonts}
\usepackage{array}
\usepackage[caption=false,font=normalsize,labelfont=sf,textfont=sf]{subfig}
\usepackage{textcomp}
\usepackage{stfloats}
\usepackage{url}
\usepackage{verbatim}
\usepackage{graphicx}
\usepackage{cite}
\usepackage{orcidlink}
\usepackage{colortbl}
\usepackage{xcolor}
\usepackage{makecell} 

\def
\BibTeX{{\rm B\kern-.05em{\sc i\kern-.025em b}\kern-.08em
    T\kern-.1667em\lower.7ex\hbox{E}\kern-.125emX}}
                                                                                                                                                              
\usepackage[ruled, linesnumbered, commentsnumbered, longend]{algorithm2e}
\SetAlCapNameFnt{\footnotesize}
\SetAlCapFnt{\footnotesize}

\usepackage{silence}
\usepackage{soul}

\usepackage{framed}
\definecolor
{shadecolor}{RGB}{250,249,249}

\begin{document}

\title{Balancing Patient Privacy and Health Data Security: The Role of Compliance in Protected Health Information (PHI) Sharing}

\author{
\IEEEauthorblockN{Md Al Amin, Hemanth Tummala, Rushabh Shah, and Indrajit Ray}
\IEEEauthorblockA{\textit{Computer Science Department, Colorado State University, Fort Collins, Colorado, USA} \\ 
\{Alamin, Hemanth.Tummala, Rushabh.Shah2, Indrajit.Ray\}@colostate.edu}
}

\maketitle

\begin{abstract}
     Protected Health Information (PHI) sharing significantly enhances patient care quality and coordination, contributing to more accurate diagnoses, efficient treatment plans, and a comprehensive understanding of patient history.  Compliance with strict privacy and security policies, such as those required by laws like HIPAA, is critical to protect PHI. Blockchain technology, which offers a decentralized and tamper-evident ledger system, hold promise in policy compliance. This system ensures the authenticity and integrity of PHI while facilitating patient consent management. In this work, we propose a blockchain technology that integrates smart contracts to partially automate consent-related processes and ensuring that PHI access and sharing follow patient preferences and legal requirements. 
\end{abstract}


\section{Introduction}   \label{sec:introduction}
Electronic health record (EHR) systems have significantly improved healthcare services, such as enhanced collaboration among healthcare professionals, more accurate diagnoses, faster treatment, and convenient access to patient-protected health information \cite{kruse2017impact}. EHR systems greatly facilitate the access and sharing of digitized healthcare information, allowing providers to easily exchange sensitive medical data with other professionals. Data sharing is essential for numerous aspects of patient care, including enhancing diagnosis and treatment plans through consultations with specialists, leveraging advanced technologies for more precise radiology and pathology analyses and diagnosis, elevating the overall quality of patient care, and others \cite{kalkman2022patients}. Furthermore, there are instances where healthcare data is utilized for research and marketing endeavors, provided specific requirements are fulfilled \cite{haddow2011nothing}. Health records can be shared through the EHR system using health information exchanges (HIE), specialized networks that rely on interoperable systems to share electronic health information seamlessly and securely \cite{li2021protocol}. Providers also share PHI through email or other electronic mediums \cite{lustgarten2020digital}. Regardless of the PHI sharing mechanism, ensuring health data security and patient privacy is mandatory.

Acquiring patient consent for healthcare information sharing is paramount for adhering to policy compliance, particularly concerning regulations like the Health Insurance Portability and Accountability Act (HIPAA) in the U.S. and the General Data Protection Regulation (GDPR) in the E.U \cite{hutchings2021systematic}. These regulatory frameworks emphasize protecting health information and upholding the patient's right to privacy. Patient consent is a cornerstone of these regulations, ensuring individuals have control over their health data and its dissemination. Under HIPAA, healthcare entities must obtain explicit consent before sharing healthcare data for purposes beyond treatment, payment, or healthcare operations. Similarly, GDPR enforces strict guidelines on data consent, processing, and privacy, offering individuals the 'right to be forgotten' and the autonomy to decide how their data is used and shared. From a policy compliance perspective, proper patient consent acquisition is a legal requirement and a trust-building measure, reinforcing the patient-provider relationship. It ensures transparency in data handling and builds patient confidence, knowing their sensitive information is shared respectfully and responsibly. As healthcare continues to integrate with various technologies, upholding these consent protocols is crucial for maintaining the security and privacy of patient data and adhering to global data protection standards.

Unauthorized health data access and disclosure are common events in healthcare industries that increase security and privacy concerns. Table \ref{table:ocr-hhs-compliance-complain} shows the number of compliance complaints received by the U.S. Department of Health and Human Services (HHS) Office for Civil Rights (OCR) \cite{rights_ocr_hipaa_2008}. The primary reasons for the complaints are (i) impermissible uses and disclosures of PHI, (ii) lack of safeguards of PHI, (iii) lack of patient access to their PHI, (iv) lack of administrative safeguards of electronic PHI, and (v) use or disclosure of more than the minimum necessary PHI. These issues can be minimized by enforcing patients' consent for data access and sharing decisions and employing proper data protection mechanisms like encryption and anonymity. Consent lets patients control their healthcare journey, enabling them to make choices that align with their best interests and well-being \cite{timmermans2020engaged}.

\begin{table}[htb]
\centering
\rowcolors{2}{purple!25}{gray!20}
\caption{OCR HHS- Compliance Complaints.} \label{table:ocr-hhs-compliance-complain}
\footnotesize
\begin{tabular}{|c| c| c| c|c|} 
\hline
  \rowcolor{purple!50}  Year &  Complains & Compliance Reviews & Technical Assistance & Total   \\
\hline
     2018 & 25089 & 438 & 7243 & 32770 \\ 
 \hline 
    2019 & 29853 & 338 & 9060 & 39251 \\
\hline
    2020 & 26530 & 566 & 5193 & 32289 \\
\hline
    2021 & 26420 & 573 & 4244 & 31237 \\
\hline
\end{tabular}
\end{table}

Enhanced security and privacy technologies are essential for protecting patient data from being compromised, misused, or disclosed. However, substantial evidence indicates that the root of many unauthorized EHR access and sharing lies in inadequate policy adoption, implementation, and enforcement \cite{lopez2023comprehensive,aljabri2022testing}. Often, users are granted access privileges inappropriately, whether intentionally or not. Policy compliance frequently falls short, and access control measures are not rigorously monitored or executed on time. A common oversight is the blanket assignment of identical roles and privileges to all employees, neglecting the nuances of individual patient-level policies. Moreover, auditing and monitoring practices are typically reactive, triggered only by serious complaints or legal mandates, rather than proactive and consistent. These policy specification and enforcement flaws significantly impact informed consent policies, underscoring the need for a more accurate and systematic approach to effectively protecting patient healthcare data and preserving privacy.

It is essential to address the following concerns to guarantee compliance with the applicable privacy and security policies, industry best practices, and contractual obligations for sharing PHI: (i) Patient-level policies or consents are often not properly or timely enforced in healthcare data sharing. (ii) Patients lack assurance that consent for access or sharing purposes is carried out strictly by designated users, and only if the stipulated conditions are met are all other requests rejected. (iii) Data sharing over email or other mediums is insecure due to the absence of encryption or the use of inadequate and weak encryption algorithms and key sizes. (iv) The centralized hospital system serves as a singular source of truth and a potential single point of failure for managing audit trails. (v) The absence of a verifiable, unaltered record for consent execution and sharing PHI highlights the need for comprehensive consent provenance. (vi) Compliance assessments and audits are not conducted accurately and timely to check compliance status.

To address the aforementioned challenges and requirements, this paper proposes a framework based on blockchain and smart contracts for managing and enforcing informed consent when sharing PHI with entities outside the treatment team. The approach ensures that PHI sharing occurs only when the sender has obtained the necessary consent from the patient and the sharing aligns with specific, predefined purposes. In addition to enforcing patient consent, this approach integrates other relevant security policies and industry best practices to ensure data protection. The HIPAA Security Rule mandates the requirements for transmission security are outlined under 45 CFR § 164.312(e)(1) Technical Safeguards \cite{chung2006overview}. However, the proposed approach does not directly guarantee security mechanisms like encryption for data protection. Instead, it leverages an honest broker who acts as a blind and secure entity to evaluate the intended PHI and certify its status as required protection mechanisms are satisfied or not \cite{alarcon2021trust}. The broker's attestation is then recorded in blockchain-based audit trails with other relevant activity data to support future compliance evaluations and validation. It supports using audit trails or provenance mechanisms based on blockchain, which is essential for keeping track of PHI-sharing activities. Moreover, the proposed framework provides a compliance-checking mechanism in data-sharing activities, ensuring adherence to applicable policies.

Smart contracts, \cite{buterin2014next}, offer an automated, transparent system that upholds the integrity and accountability of the consent for sharing PHI. Through this smart contract-based approach, the proposed framework not only automates processes but also guarantees the accurate execution of informed consent, thereby enhancing the security and reliability of PHI sharing. Blockchain technology ensures the immutability of submitted records, safeguarding the integrity of the audit trail and enabling the detection of any unauthorized alterations. Blockchain security features, including non-repudiation, ensure that participants cannot deny their actions \cite{le2021systematic}.

This work is the first to capture patients' informed consent for PHI sharing to ensure policy compliance through preserving provenance and conducting compliance checking. It also considers and enforces other applicable security policies and industry best practices mandated by the various laws, regulations, standards, and contractual obligations to meet the compliance requirements. Significant contributions include (i) implementing a mechanism to capture patients' consent for sharing healthcare data beyond the treatment team members. (ii) Storing obtained consents in decentralized and distributed networks (blockchain) to overcome a single point of truth sources and failure. (iii) Considering applicable security and privacy policies, regulatory requirements, and contractual obligations to ensure compliance-based sharing. (iv) Enforcing informed consent and applicable policies while making authorization decisions to share health records. (v) Equipping blockchain-based audit trail mechanisms to guarantee data provenance. (vi) Incorporating compliance assessment methods to identify compliance and non-compliance PHI sharing. (vii) Offering consent services to provide precise and comprehensive insights into the consent granted and the extent of its execution.

The remainder of the paper is organized as follows: Section \ref{sec:related-works} discusses some works that are related to this work. The proposed approach is explained in Section \ref{sec:proposed-approach} with the necessary components. Section \ref{sec:provenance-for-compliance} gives the structure of audit trails and consent provenance. The compliance verification mechanism is explained in Section \ref{sec:policy-compliance-verification}. Section \ref{sec:consent-services} discusses essential services for given executed consents. The experimental evaluations of the proposed approach are provided in Section \ref{sec:experimental-proposition} for PPA integrity storage, patient contracts deployment, consent storage cost, and writing and reading time.  Section \ref{sec:conclusion-future-directions} concludes the paper with future research directions on consent management.

\section{Related Work} \label{sec:related-works}
Blockchain technology has increasingly been adopted in healthcare for various services, particularly for sharing protected health information among healthcare providers, patients, and other stakeholders. Blockchain facilitates a more efficient, transparent, and patient-centered delivery of healthcare services, making it an essential component in modern healthcare infrastructure. Fan et al., \cite{fan2018medblock}, proposed a blockchain-based secure system, MedBlock, to share electronic medical records among authorized users. It provides security and privacy with access control protocols and encryption technology while sharing patient healthcare data.

Shah et al., \cite{shah2019crowdmed}, proposed a medical data management framework to facilitate data sharing. It gives patients full control over access to their medical data. It also ensures that patients know who can access their data and how it is used. Zhuang et al., \cite{zhuang2020patient}, addressed a blockchain-based patient-centric health information-sharing mechanism protecting data security and privacy, ensuring data provenance, and providing patients full control over their health data. However, consent structure and compliance requirements are not addressed, which are very important to give patients confidence in how their consent is executed and how data is protected.

Alhajri et al., \cite{alhajri2022privacy}, explored the criticality of implementing legal frameworks to safeguard privacy within fitness apps. By examining how various fitness apps handle consent and privacy policies, their research highlighted the crucial role of consent as outlined in the GDPR. The authors proposed the adoption of blockchain technology as a means to govern user consent for sharing, collecting, and processing fitness data, ensuring a process centered around human needs and compliant with legal standards. Nonetheless, the study failed to present a technical architecture for their blockchain-based proposal.

Amofa et al. approached a blockchain-based personal health data sharing framework with an underlying mechanism to monitor and enforce acceptable use policies attached to patient data \cite{amofa2018blockchain}. Generated policies are consulted with smart contracts to make decisions on when the intended data can be shared or otherwise. All entities cooperate to protect patient health records from unauthorized access and computations. Balistri et al., \cite{balistri2021blockhealth}, designed the \textit{BlockHealth} solution for sharing health data with tamper-proofing and protection guarantees. They store the patient's healthcare data in a private database, and the hash of the healthcare data is stored in the blockchain to ensure data integrity. Shen et al., \cite{shen2019medchain}, proposed \textit{MedChain}, a blockchain-based health data sharing approach where data streams are continuously generated from sensors and other monitoring devices from various patients' bodies. The collected data are shared with laboratories and health organizations for diagnosis, advanced treatment, and further research.

The above-mentioned papers summarized the application and benefits of using blockchain for healthcare data sharing and essential services. However, they failed to address the security and privacy requirements mandated by various laws and regulatory agencies, such as HIPAA and GDPR. The major requirements demand patient consent and proper protection, such as encryption, while sharing health records. In addition, it is crucial to maintain audit logs and check that those activities did not violate any policies. This paper proposes sharing informed consent as the smart contract for authorization with provenance and compliance-checking mechanisms.

\section{Proposed Approach} \label{sec:proposed-approach}

The main objective is to ensure compliance with applicable security and privacy policy for PHI sharing. To ensure compliance, we need proper policy enforcement, including maintaining provenance and performing compliance status checks promptly and properly. For enforcement, this paper considers patient-informed consent, where the sender has permission from the patient to share the intended PHI with the receiver for specific purposes. Also, proper data protection mechanisms are considered. However, instead of ensuring data protection directly, this work leverages an honest broker to verify and certify the data protection mechanism. PHI-sharing activities are recorded as audit trails to provide provenance and reconstruct events in a manner that reflects their actual occurrence. A private blockchain-based approach is proposed (Section \ref{sec:provenance-for-compliance}). Finally,  a blockchain consensus mechanism called Proof of Compliance (PoC) is approached, Section \ref{sec:policy-compliance-verification}, for performing auditing. This audit rigorously examines the enforcement actions against the policy standards and informed consent, using the provenance data to verify and certify the policy's compliance status while sharing health records. The seamless connection between policy enforcement, provenance, and the auditing process forms the backbone of a secure and compliant system.

\subsection{Patient-Provider Agreement (PPA)}
\begin{figure}[htb]
    \centering
    \includegraphics[width=\linewidth]{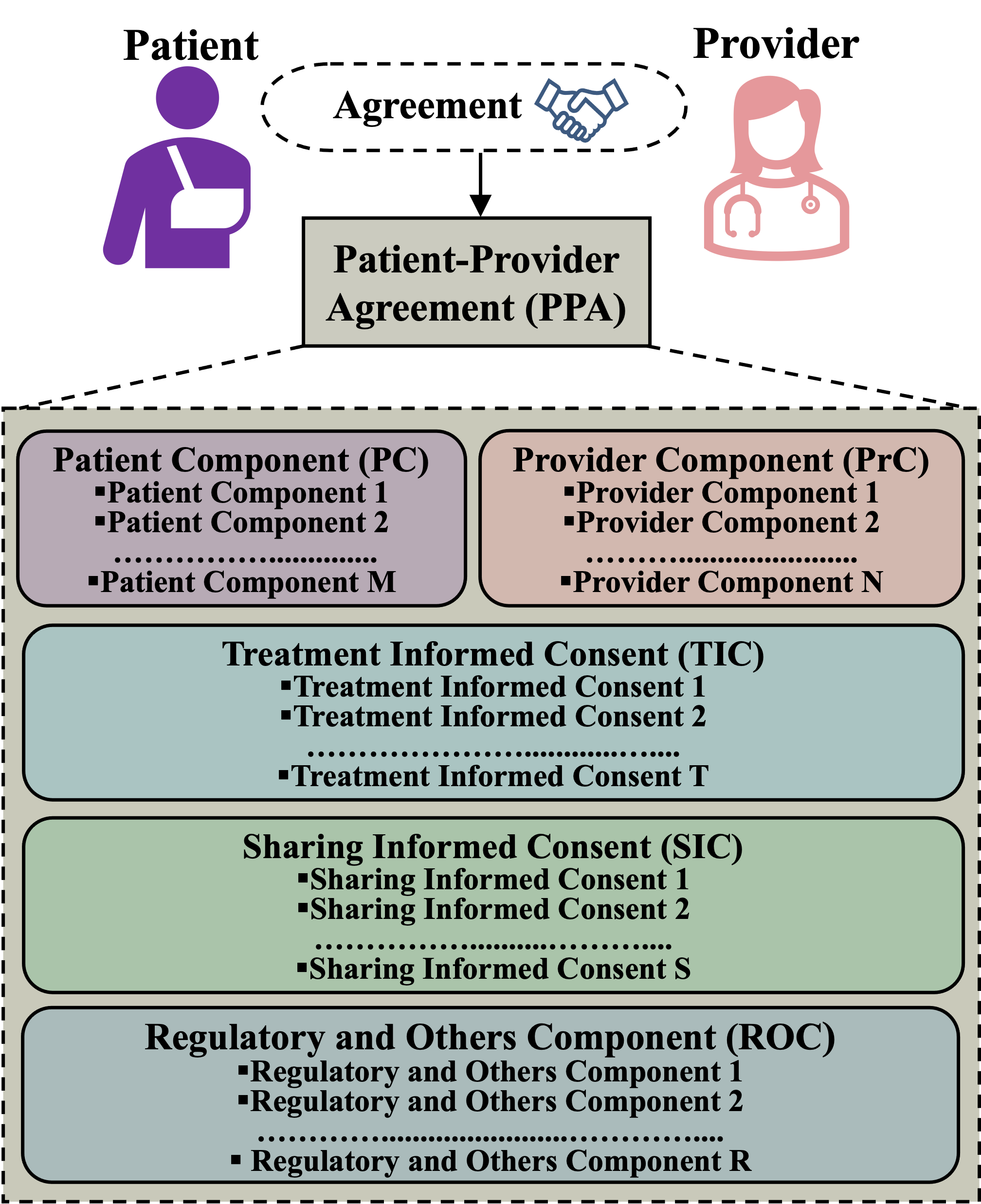}
    \caption{Patient-Provider Agreement (PPA) Components.} \label{fig:patient-provider-agreement}
\end{figure}

The patient-provider agreement, or PPA, aims to determine who is responsible for what in treatment. A PPA is formed when a patient visits a hospital and is properly documented to deliver healthcare services. It differs from organization to organization. Healthcare organizations adjust what they need from patients and what they expect from them to match those needs, treatments, and responsibilities. This is done based on the nature and needs of treatment and services. Also, the components and representation of the PPA depend on the hospital or clinic. Figure \ref{fig:patient-provider-agreement} shows the structure of a PPA, and Algorithm \ref{alg:patient-provider-agreement} illustrates the gradual processes for creating a PPA with the required components. The main concept of PPA is adopted from \cite{al2023informed}. The authors focused on consent management for medical treatment and diagnosis purposes, mainly for the treatment team members. They did not include patient consent and other requirements for health data sharing beyond the treatment team. This paper extends the PPA structure to analyze the requirements and formalize the consent components for PHI sharing. A PPA is formally composed of five tuples: \[ PPA = (PC, PrC, TIC, SIC, ROC) \] satisfying the following requirements:
\begin{itemize}
    \item[(A)] $PC$ is a finite set of patient components containing the patient's personal information, contact information, mailing information, pharmacy information, billing and insurance information, emergency contact, and others. The patient is responsible for providing and maintaining these components' valid, accurate, and updated information.
    
    \item[(B)]  $PrC$ is a finite set of provider components, including the treatment team, prescription, and others. The provider is responsible for creating an effective team to provide appropriate care. Everything from treatment to insurance coverage and billing is considered during the patient treatment period.
    
    \item[(C)] $TIC$ is a finite set of treatment informed consent components. It denotes that the patient has permitted the designated treatment team to access medical records. Treatment team members include doctors, nurses, support staff, lab technicians, billing officers, emergency contact persons, and others assigned by the authority.  Some outsider members are insurance agents, pharmacists/pharmacy technicians, doctors/lab technicians from another hospital, etc.
   
    \item[(D)] $SIC$ is a finite set of sharing informed consent components. It denotes the patient's consent to sharing medical data for a specific purpose. Both the sender and the receiver must have consent. The primary purpose of this work is $SIC$, including (i) identifying, capturing, and storing consent components, (ii) enforcing consents with other applicable security policies and industry best practices to ensure policy compliance while making PHI-sharing decisions, (iii) defining and capturing provenance information with the enforced consents to maintain audit trails, (iv) performing compliance checking using consensus mechanisms; (v) providing services for both given and executed consents, etc. It does not consider other components: $PC$, $PrC$, $TIC$, and $ROC$.
 
    \item[(E)] $ROC$ is a finite set of regulatory and other components. It has applicable security and privacy policies to comply with the requirements of local government, state government, federal government, foreign government, and regulatory agencies (HIPAA, GDPR) if necessary. It also includes contractual obligations in some cases.
\end{itemize}

\RestyleAlgo{ruled}
\SetKwComment{Comment}{/* }{ */}

\begin{algorithm}[htb]
    \scriptsize
    \SetKwInOut{KwData}{Input}
    \SetKwInOut{KwResult}{Result}
    \DontPrintSemicolon

\caption{Patient-Provider Agreement (PPA) Formation.}\label{alg:patient-provider-agreement}
\KwData{(i) $PC$, (ii) $PrC$, (iii) $TIC$, (iv) $SIC$, (v) $ROC$, (vi) $\mathbb{R}_{PPA}$, (vii) $\mathbb{BN}_{SC}$}
                \textcolor{blue}{\Comment*[r]{$\mathbb{R}_{PPA}$: secured PPA repository, $BN_{SC}$: blockchain network smart contract}}
\KwResult{A formal $PPA$}

\textbf{Input Parameters Initialization} \vfill 
  $PPA_i \gets  \{ PC_i, PrC_i, TIC_i, SIC_i, ROC_i\}$ where $i$ is patient identity \vfill
    \quad \textit{(i)} $PC \gets \{ PC_1, PC_2, PC_3, PC_4, PC_5, PC_6......................PC_M \}$\;  \vfill
    \quad \textit{(ii)} $PrC \gets \{ PrC_1, PrC_2, PrC_3, PrC_4, PrC_5, PrC_6..........PrC_N \}$\;  \vfill
    \quad \textit{(iii)} $TIC \gets \{ TIC_1, TIC_2, TIC_3, TIC_4, TIC_5, TIC_6........TIC_T \}$\; \vfill
    \quad \textit{(iv)} $SIC \gets \{ SIC_1, SIC_2, SIC_3, SIC_4, SIC_5, SIC_6..............SIC_S \}$\; \vfill
    \quad \textit{(v)} $ROC \gets \{ ROC_1, ROC_2, ROC_3, ROC_4, ROC_5, ROC_6...ROC_R \}$\; \vfill
 
 \textbf{PPA Components Integrity Calculation} \vfill        \textcolor{blue}{\Comment*[r]{$\mathbb{H}(\partial)$ calculates hash of $\partial$}}
    \quad \textit{(a)} $\mathbb{H}_{PC}  \gets \mathbb{H} ( PC_1, PC_2, PC_3, PC_4, PC_5, PC_6....................PC_M)$\; \vfill
    \quad \textit{(b)} $\mathbb{H}_{PrC} \gets \mathbb{H} ( PrC_1, PrC_2, PrC_3, PrC_4, PrC_5, PrC_6.........PrC_N)$\;  \vfill
    \quad \textit{(c)} $\mathbb{H}_{TIC} \gets \mathbb{H} ( TIC_1, TIC_2, TIC_3, TIC_4, TIC_5, TIC_6........TIC_T)$\;  \vfill
    \quad \textit{(d)} $\mathbb{H}_{SIC} \gets \mathbb{H} ( SIC_1, SIC_2, SIC_3, SIC_4, SIC_5, SIC_6.............SIC_S)$\;  \vfill
    \quad \textit{(e)} $\mathbb{H}_{ROC} \gets \mathbb{H} ( ROC_1, ROC_2, ROC_3, ROC_4, ROC_5, ROC_6..ROC_R )$\;\vfill
    \quad \textit{(f)} $\mathbb{H}_{PPA_i} \gets  \mathbb{H}( \mathbb{H}_{PC}, \mathbb{H}_{PrC}, \mathbb{H}_{TIC}, \mathbb{H}_{SIC}, \mathbb{H}_{ROC})$\; \vfill

 \textbf{PPA Finalization} \vfill 
  \eIf{$PPA_i$ is complete}{
        \textcolor{blue}{\Comment*[r]{presence of $PC$, $PrC$, $TIC$, $SIC$, $ROC$}}
            \eIf{$(\mathbb{R}_{PPA} + PPA_i)$ contains no conflicts}{    
                   \textit{(i)} do $\mathbb{R}_{PPA} \gets (\mathbb{R}_{PPA} + PPA_i)$\;
                   \textit{(ii)} add $\mathbb{ID}_{PPA_i}$ to patient profile, $\mathbb{P}_i$\;
                   \textit{(iii)} call $\mathbb{BN}_{SC} (\mathbb{ID}_{PPA_i}, \mathbb{H}_{PPA_i})$\; 
                        \textcolor{blue}{ \Comment*[r]{PPA integrity verification reference}}
                   
                   \textbf{\textit{Return: }} Success ($PPA_i$ added to $\mathbb{R}_{PPA}$)\;
            }{
                \textbf{\textit{Error: }} $(\mathbb{R}_{PPA} + PPA_i)$ contains conflicts\;
                            \textcolor{blue}{ \Comment*[r]{ $PPA_i$ revision required to add}}
            }
   }{
        \textbf{\textit{Error: }}$PPA_i$ cannot be created (incomplete PPA)\;  
 }
\end{algorithm}

\subsection{Sharing Informed Consent (SIC)}
\begin{figure}[tb]
    \centering
    \includegraphics[width=\linewidth]{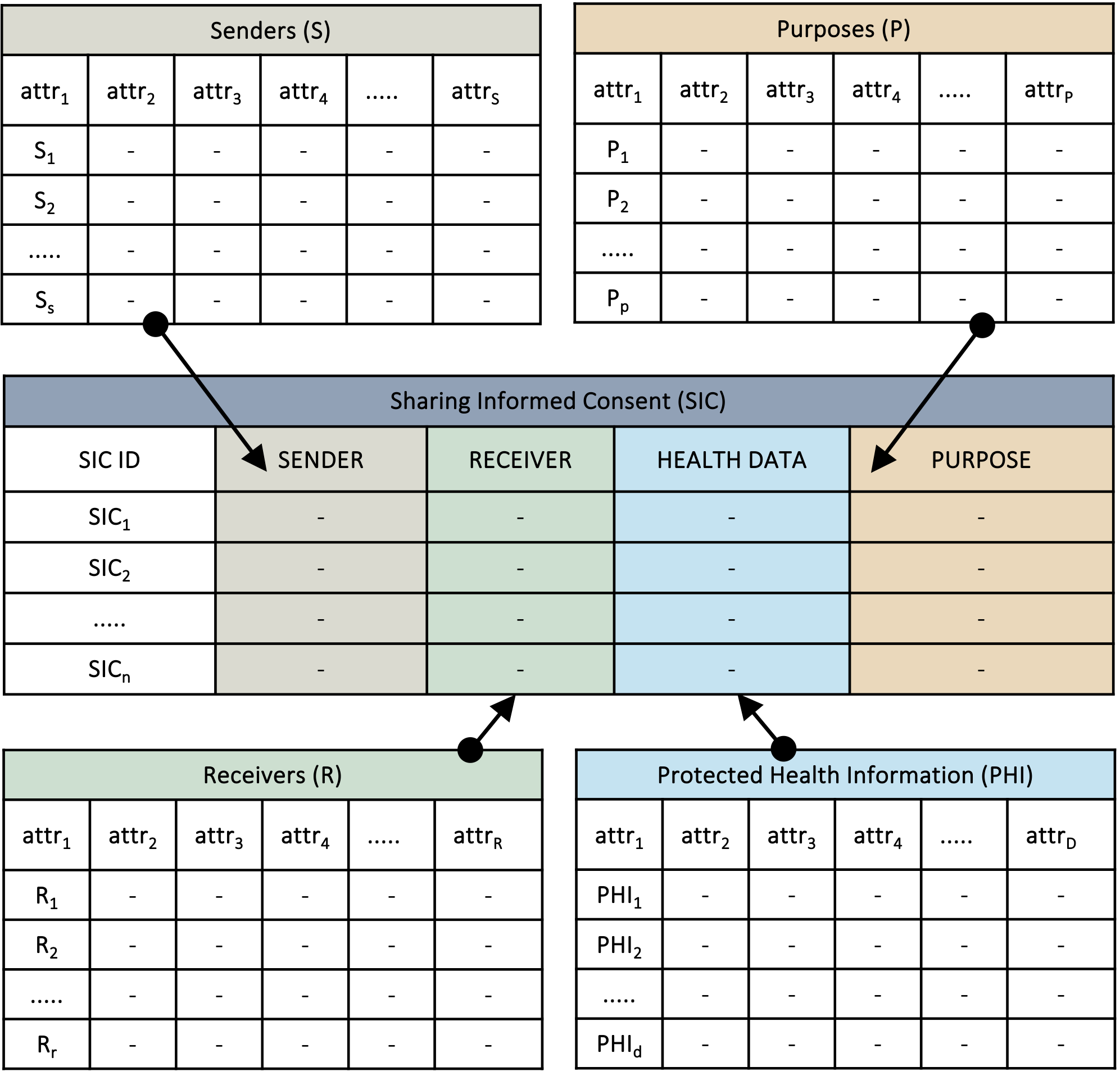}
    \caption{Sharing Informed Consent (SIC) Structure.} \label{fig:informed-consent-components}
\end{figure}

Before approving, patients need to know clearly about the sharing informed consent, particularly who can share which PHI with whom for what purposes—and also the protection mechanism while sharing PHI during transmission over the network. Figure \ref{fig:informed-consent-components} shows the SIC conceptual framework structure. Sharing informed consent is formally composed of four tuples: \[ SIC = (S, R, PHI, P) \]  satisfying the following requirements:
\begin{itemize}
    \item[(a)] $S$ is a finite set of authorized senders denoted as $\lbrace S_1, S_2, S_3, ......S_s\rbrace$ for $s$ number of senders. The sender can share certain healthcare data with the receiver, who has permission from the patient. The sender may be a member of the patient treatment team or anyone from the provider.
    
    \item[(b)]  $R$ is a finite set of authorized users who receive protected health information from authorized senders. A finite set of $r$ number authorized receivers denoted as $\lbrace R_1, R_2, R_3, ......R_r\rbrace$. The receiver may be from other hospitals, labs, medical research institutes, pharmaceutical companies, marketing departments, government officials, etc.
 
    \item[(c)] $PHI$ is a finite set, $d$ number, of health data denoted by  $\lbrace PHI_1, PHI_2, PHI_3, ......PHI_d\rbrace$. It is an electronic version of a patient's medical data that healthcare providers keep over time. They are protected health information and contain sensitive patient information. PHI must be protected from any kind of unauthorized access, disclosure, and sharing. Table \ref{table:patient-records} shows ten (10) types of PHI, considered for each patient, with PHI ID, name, description, and potential creators.

    \item[(d)] $P$ is a finite set of purposes. It indicates the objective of the PHI sharing by the senders with the receivers. Receivers must use the received PHI for the intended purposes. A finite set of purposes, a $p$ number, can be denoted as $\lbrace P_1, P_2, P_3, ......P_p\rbrace$.
\end{itemize}

\begin{table*}[htb]
    \centering
        \rowcolors{2}{teal!25}{gray!20}
        \caption{Sample Patient Protected Health Information (PHI) Structure.} \label{table:patient-records}
        \scriptsize
    \begin{tabular}{|c| l| l| l|} 
        \hline
    \rowcolor{teal!60}  \textbf{PHI ID} &  \textbf{PHI Name} & \textbf{PHI Description}  & \textbf{PHI Creator}    \\
        \hline
     PHI-1001 & Demographic Information & Basic personal information like name, date of birth, gender, contact & Patient, Support Staff\\ 
         \hline 
     PHI-1002 &  Previous Medical History & Old medical records from another hospitals and providers & Patient, Support Staff\\
        \hline
     PHI-1003 & Immunizations, Vaccinations  & Immunization records that are administered over time & Patient, Pathology Lab Technician \\
        \hline
     PHI-1004& Allergies & Various allergies sources, triggering condition, remediation & Patient, Support Staff, Path Lab Tech\\
        \hline
     PHI-1005 & Visit Notes & Physiological data,  advises, follow-up, visit details & Doctor, Nurse \\
        \hline
     PHI-1006& Medications, Prescription &Pharmacy information, prescribed medications like name, dosage & Doctor \\
        \hline
     PHI-1007 & Pathology Lab Works & Biological samples analysis like blood, tissue, other substances & Pathology Lab Technician \\
        \hline
     PHI-1008 & Radiology Lab Works & Imaging results such as X-rays, CT, MRI, Ultrasound, PET scans & Radiology Lab Technician \\
        \hline
     PHI-1009 & Billing, Insurance & Bank account, credit/debit card,  and insurance policy information & Patient, Support Staff, Billing Officer \\
        \hline
     PHI-1010 & Payer Transactions & Bills of doctor visit, lab works, and medications & Billing Officers, Insurance Agent \\
        \hline
    \end{tabular}
\end{table*}

The objective of sharing protected health information outlines the specific reasons for its sharing. The recipient must utilize the shared PHI exclusively for its designated purpose. The potential reasons for sharing PHI in this study include, but are not limited to:

\begin{itemize}
    \item[(i)] \textbf{\textit{Treatment:}} Providers or patients need to share PHI with other providers from external hospitals to provide better treatment. Also, patients must move to different regions, like states or countries, due to family movement, job transfers, or new jobs. Patients need to share or transfer healthcare data from the previous providers to the current.
    
    \item[(ii)] \textbf{\textit{Diagnosis:}} Present providers sometimes need more skilled human resources, appropriate machinery, instruments, or sophisticated technology to diagnose disease. But it is urgently required to do that to give proper treatment and services to save patients' lives or minimize damages. Patients' health data must be transferred or shared with other providers or labs to complete diagnosis and make proper treatment plans for the patients.

    \item[(iii)] \textbf{\textit{Marketing:}} Healthcare data sharing for marketing purposes involves using patient data to promote healthcare services, products, or initiatives. This can help healthcare providers tailor their services to patient needs, inform patients about new treatments or products, and improve patient engagement. Only the receiver entity can use the shared data as intended and should not share it with other associates for extended business purposes.
   
    \item[(iv)] \textbf{\textit{Research:}} Sharing PHI for medical research purposes holds significant potential for advancing medical knowledge, leading to breakthroughs in understanding diseases, improving and developing new treatments, improving healthcare systems and services, and enhancing patient outcomes. Patients' privacy and rights must be respected.
\end{itemize}

Other purposes might exist depending on the nature and requirements of the treatment, patient conditions, provider business policy, etc. This study considers only the four purposes mentioned above. After receiving shared data, the receiver performs specified operations to complete the job. It is assumed that the receiver cannot share data with other users who do not have permission from the patients. More specifically, the receiver's healthcare system does not allow the sharing of PHI by any means, like printouts, email, or screenshots. However, this paper doesn't provide detailed mechanisms or techniques for preventing data sharing without patients' consent at the receiver end.

\subsection{SIC Smart Contract Deployment}

Once a Patient-Provider Agreement, or PPA, is created and stored in the repository, all sharing informed consent components are deployed to the blockchain network. For each patient, there is one smart contract that contains all consents for that particular patient. If there isn't a smart contract, the authority deploys one, transfers ownership to the patient, and updates the contract address to the patient's profile and hospital systems. The contract address is an identifier for a smart contract in the blockchain network. This smart contract-based approach provides an automated system and guarantees the integrity and accountability of the deployed consents. Once consents are deployed or added to the smart contract, they cannot be altered. The authorization module needs to access these smart contracts to make decisions considering the sender, receiver, purpose attributes, environmental factors, organizational policies, regulatory frameworks, etc.

\begin{figure}[tb]
    \centering
    \includegraphics[width=\linewidth]{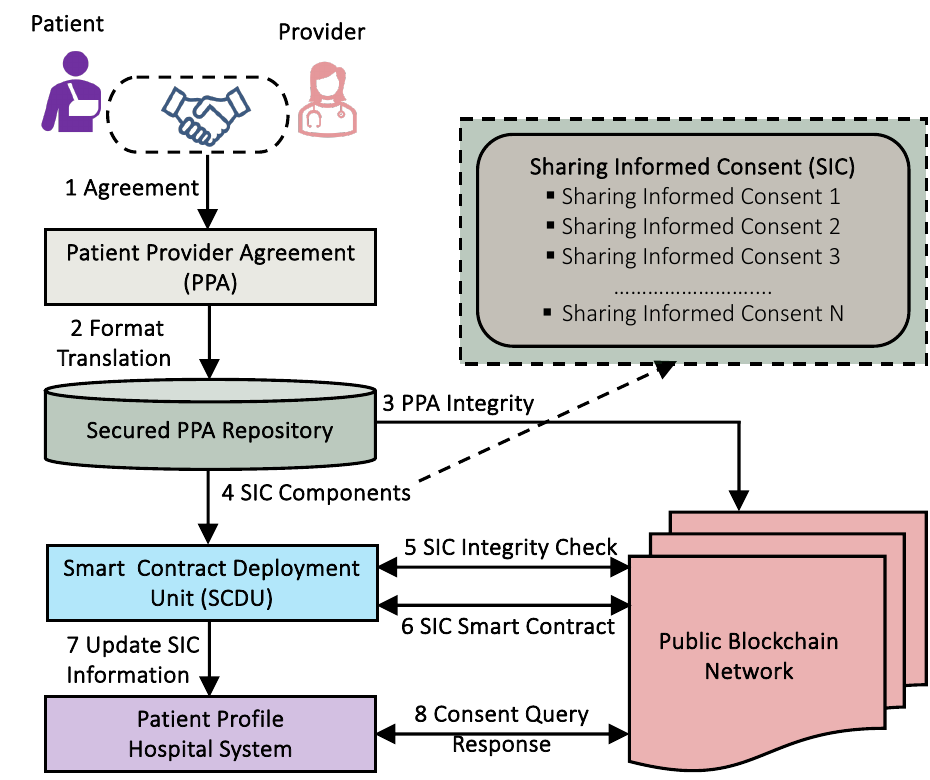}
    \caption{SIC Smart Contract Deployment Process.} \label{fig:consent-smart-contract-deployement}
\end{figure}

Upon finalizing the PPA, it transforms and secures storage in a PPA repository. Subsequently, an integrity marker, such as a hash ($\mathbb{H}_{PPA_i}$) generated by the Algorithm \ref{alg:patient-provider-agreement}, is stored on the blockchain alongside the PPA ID for later modification detection. These are depicted in Steps 2 and 3 in Figure \ref{fig:consent-smart-contract-deployement}. The Smart Contract Deployment Unit (SCDU) then gathers all components of the informed consent from the PPA (Step 4). It verifies their integrity to ensure no deliberate or accidental alterations have occurred (Step 5). As a secure entity, the SCDU does not alter consent components, noting that any modification invalidates the consent. If the consents remain unmodified, the SCDU creates and deploys the corresponding smart contracts on the blockchain network (Step 6) and then updates the patient's profile and the hospital system (Step 7). Users can make queries with the required credentials regarding informed consent and get responses in Step 8 from the blockchain network.

\subsection{Honest Broker, Applicable Policies and Industry Best Practices}

Alongside patient consent, the proposed approach incorporates relevant security policies and industry best practices before sharing protected health information. For instance, a security policy might require a data protection mechanism during data transfer between systems. For treatment and diagnosis purposes, encryption is a recommended protection method. 

As an industry best practice, the Advanced Encryption Standard (AES) is preferred over the Data Encryption Standard (DES). Furthermore, it advises using a robust, lengthy encryption key (256 bits) rather than a weaker, shorter one (64 or 128 bits). The sender must encrypt the intended PHI using the AES-256 algorithm while leaving it in the system for treatment and diagnosis. However, this proposed approach does not encrypt the healthcare data directly or ensure a strong key size while encrypting the intended healthcare data. Also, it does not address the key management mechanisms such as creation, storage, sharing, updating, deleting, etc. It is assumed that the key management is done securely and separately.

Similarly, anonymity is a recommended protection method for marketing and research purposes, where patient identifiers must be removed before sharing. The targeted PHI must be anonymous using proper techniques and tools before sending the data from the host healthcare system to the receiver. The host system indicates where patients' PHI is created or presently stored. Healthcare organizations deploy appropriate encryption and anonymity mechanisms. This study does not directly ensure PHI encryption and anonymity. Instead, this approach leverages an honest broker, a trusted entity that evaluates the encryption algorithm, key size, and data anonymity status \cite{alarcon2021trust}. After checking, the honest broker certifies or attests to the status, which is recorded in audit trails as proof for policy compliance verification, along with other components like sharing informed consents, timestamps, etc. 

Depending on the specific policies and practices of the healthcare organization, this broker could be either a human or a non-human (automated) entity. The honest broker's role is confined; it does not share healthcare data with other entities. It also does not analyze data to gain insights about the patient or share those insights. Effectively, it functions as a 'blind' entity, ensuring encryption standards and the anonymity status of the PHI without engaging with the actual data content.

\subsection{PHI Sharing Authorization Process}

\begin{figure*}[tb]
    \centering
    \includegraphics[width=\linewidth]{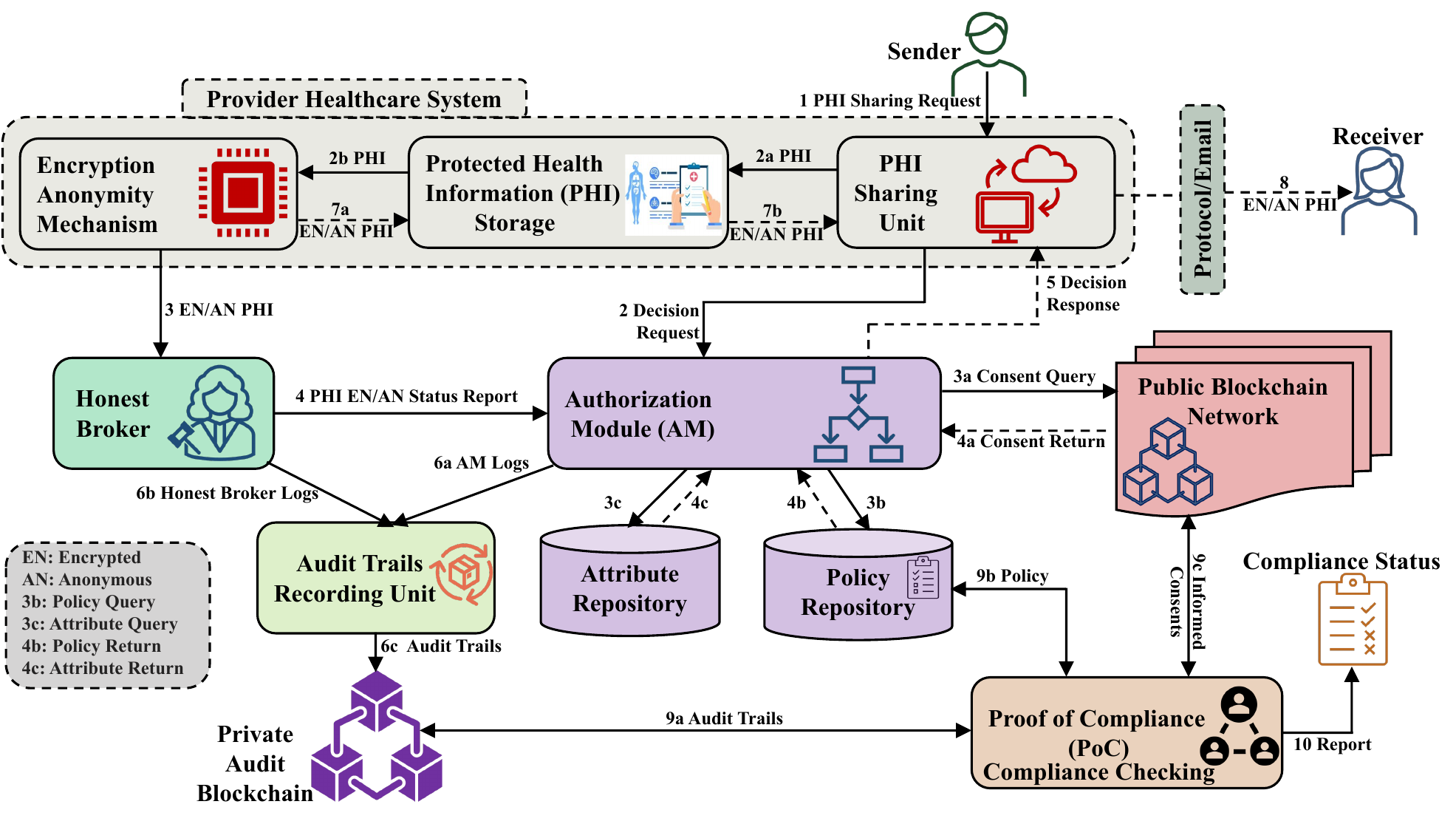}
    \caption{Compliance-Based PHI Sharing Authorization Process.} \label{fig:phi-sharing-authorization-process}
\end{figure*}

Consent enforcement ensures that related consents are executed while making decisions for the PHI sharing requests. All consents are stored on the public blockchain network as smart contracts and cannot be enforced until they are called. The authorization module (AM) considers sharing informed consent with applicable policy and required attributes while making decisions. The attributes may be subject, object, operation, and environmental attributes. The sender must provide the necessary credentials for identification and authentication. Figure \ref{fig:phi-sharing-authorization-process} shows the informed consent enforcement for PHI-sharing authorization. 

A sender submits a data sharing request to the PHI sharing unit in \textit{Step 1}. Sharing unit forwards request to authorization module for decision in \textit{Step 2}. It also requests that the PHI storage unit send the intended PHI to the protection mechanism unit in \textit{Steps 2a} and \textit{2b}. The honest broker receives encrypted or anonymized data in \textit{Step 3}. After analyzing, it sends a report to AM in \textit{Step 4}. The AM queries the blockchain network through the corresponding smart contract to get sharing informed consent information for the sharing request in \textit{Step 3a} and \textit{4a}. It also makes queries for requests related to applicable policies and required attributes in \textit{Steps 3b} and \textit{3c}. It receives the policy and attributes in \textit{Steps 4b} and \textit{4c}. After evaluating, it makes an authorization decision and sends it to the sharing unit in \textit{Step 5}. If the request is approved, the sharing unit gets encrypted or anonymized data based on the purpose in \textit{Steps 7a} and \textit{7b}. Then, it delivers the intended PHI through email or protocol to the receiver in \textit{Step 8}. 

The audit trail recording unit collects logs from AM in \textit{Step 6a} and from the honest broker in \textit{Step 6b}. It combines logs and stores as an audit trail in \textit{Step 6c} in Private Audit Blockchain. Section \ref{sec:provenance-for-compliance} discusses block structure and others. The compliance status checking is done in \textit{Steps 9a, 9b}, and \textit{9c} by the Proof of Compliance consensus mechanism. Compliance status reports are produced in \textit{Step 10}. Section \ref{sec:policy-compliance-verification} discusses the required mechanism. For this study, it is considered that the authorization module is not compromised or tampered with. It is the reference monitor for making access decisions and must be tamper-proof \cite{mulamba2017resilient}. Also, the communication channel between AU and the smart contract access points or apps is secured from malicious users.

\section{PHI Sharing Provenance} \label{sec:provenance-for-compliance}
Enforcing an applicable set of policies is crucial, but preserving data provenance to show adherence to these policies is also essential. Nevertheless, policy compliance cannot be quantified or confirmed in isolation. An independent auditor conducts a thorough policy audit to verify compliance with the policy, utilizing the available provenance data to ascertain and certify the policy's compliance status. For an accurate policy compliance assessment, two critical elements must be diligently maintained: \textit{(i) consent and policy lineage} and \textit{(ii) PHI sharing activity audit trails}. This section contains the detailed provenance mechanisms dedicated to preserving the policy lineage's integrity and ensuring the audit trails' authenticity.

\subsection{Consent and Policy Lineage}
Policy lineage involves a comprehensive record of all policies that guide the authorization module's decisions. It's a transparent and traceable record of the policy history and its application in decision-making processes. For this study, sharing informed consent is mainly considered for decision-making. Since all consents are deployed as smart contracts, blockchain networks can create policy lineages. However, this paper does not consider other HIPAA-related policies, such as physical security, provider training, etc \cite{chung2006overview}.

\subsection{PHI Sharing Activity Audit Trails}
Integrity in policy enforcement ensures that events are documented faithfully, reflecting the sequence and nature of actions taken. This authenticity is crucial for transparency and accountability. Provenance plays a key role by offering a detailed and unalterable history of policy enforcement actions as they are carried out, safeguarding against any tampering of records. The alteration of audit trails or unauthorized access to healthcare data is strictly prohibited to maintain the sanctity of the process. Maintaining the integrity of the audit trail is essential for policy compliance assurance. If integrity is compromised, checking compliance status to find compliance and non-compliance cases is questionable. The blockchain provides these requirements as ledger properties. This work adopts private blockchain as an audit trail storage system.

\begin{figure}[tb]
    \centering
    \includegraphics[width=\linewidth]{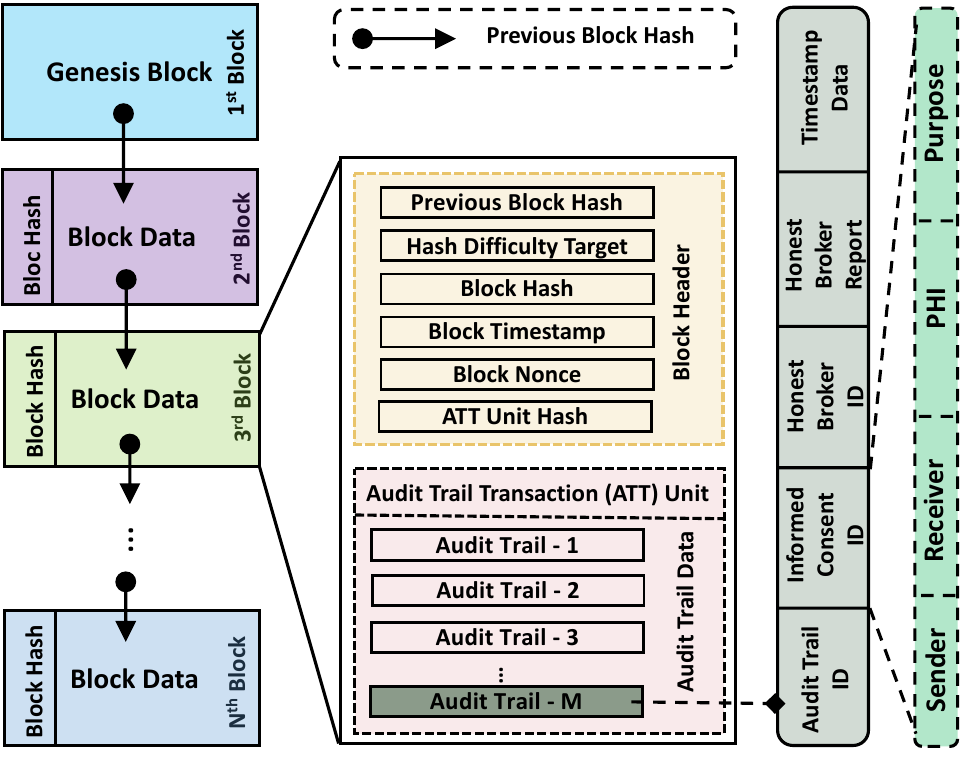}
    \caption{Audit Blockchain Block Structure.} \label{fig:audit-block-structure}
\end{figure}

Figure \ref{fig:audit-block-structure} illustrates the private audit blockchain's block components and structure.  Each block has a block header part that contains block metadata and a data part that stores the audit trail data. Each audit trail has five components: \textit{(i) audit trail ID; (ii) informed consent ID or SIC ID; (iii) honest broker ID; (iv) honest broker report;} and \textit{(v) timestamp data.} The audit trail ID provides unique identifiers; the informed consent ID, or SIC ID, indicates the consent that is executed to share the intended PHI. From SIC ID, it is possible to get the components: sender, receiver, PHI, and purpose. The honest broker ID indicates which broker certifies or attests to the intended PHI's protection status (encryption or anonymity). Finally, the timestamp means the time when the sharing authorization is done. Steps \textit{6a, 6b,} and \textit{6c} in Figure \ref{fig:phi-sharing-authorization-process} show the process of capturing audit trails from the authorization module and honest broker.

\begin{figure}[tb]
    \centering
    \includegraphics[width=\linewidth]{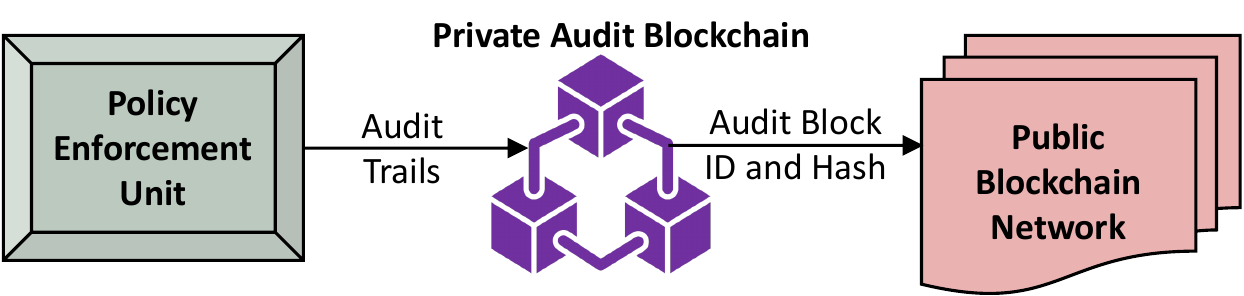}
    \caption{Storing Audit Blockchain Block ID and Hash.} \label{fig:private-public-blockchain}
\end{figure}

Enforcement activity data is collected and stored in a private blockchain known as an audit blockchain as immutable records to ensure consent provenance and maintain compliance. The private blockchain network is managed and maintained by an authority, which means reading and writing permissions are given to limited participants or users. In this case, the trust and transparency of the private blockchain are questionable. It doesn't provide a public eye to maintain trust and transparency. Storing audit trails on the public blockchain gives trust and transparency, which is another issue to consider. Firstly, audit trails contain sensitive information like user activities, and storing them on a public blockchain creates security and privacy concerns. Secondly, audit trails produce enormous amounts of data, which requires a lot of money to store on the public blockchain. This is not feasible from a business perspective, as it increases business operation, treatment costs, and service charges. 

To overcome the aforementioned issues, this research stores audit trail data on a private blockchain called the private audit blockchain. Then, it stores the private audit blockchain block ID and hash as integrity on the public blockchain. Storing block ID and integrity requires a small cost and provides trust and transparency. Any modifications to private audit blockchain data can be detected by comparing the block's current and stored hashes with those on the public blockchain. Figure \ref{fig:private-public-blockchain} shows the private and public blockchain relationship for storing audit block ID and integrity in a public blockchain like Ethereum. We have configured a private blockchain that is based on the Ethereum client \cite{samuel2021choice} with the necessary smart contracts and API for capturing and storing audit trail data in the audit blockchain.

\section{Compliance Verification} \label{sec:policy-compliance-verification}
Enforcing applicable policies and maintaining audit trails are insufficient to ensure policy compliance. There must be some mechanism to check compliance status using deployed and enforced policies with audit trails. The compliance checker must be an independent and separate entity from the policy enforcer and audit trail unit. This paper proposes a blockchain consensus mechanism to perform compliance-checking operations on the audit trails using deployed sharing informed consents (SIC) and other applicable policies. The consensus mechanism, called Proof of Compliance (PoC), is governed by a set of independent, distributed, and decentralized auditor nodes. Section \ref{sec:proposed-approach} discusses the sharing informed consent structure and deployment process as the smart contract in the public blockchain. Section \ref{sec:provenance-for-compliance} gives the audit trail capturing and storing mechanism. 

\begin{figure}[htb]
    \centering
    \includegraphics[width=\linewidth]{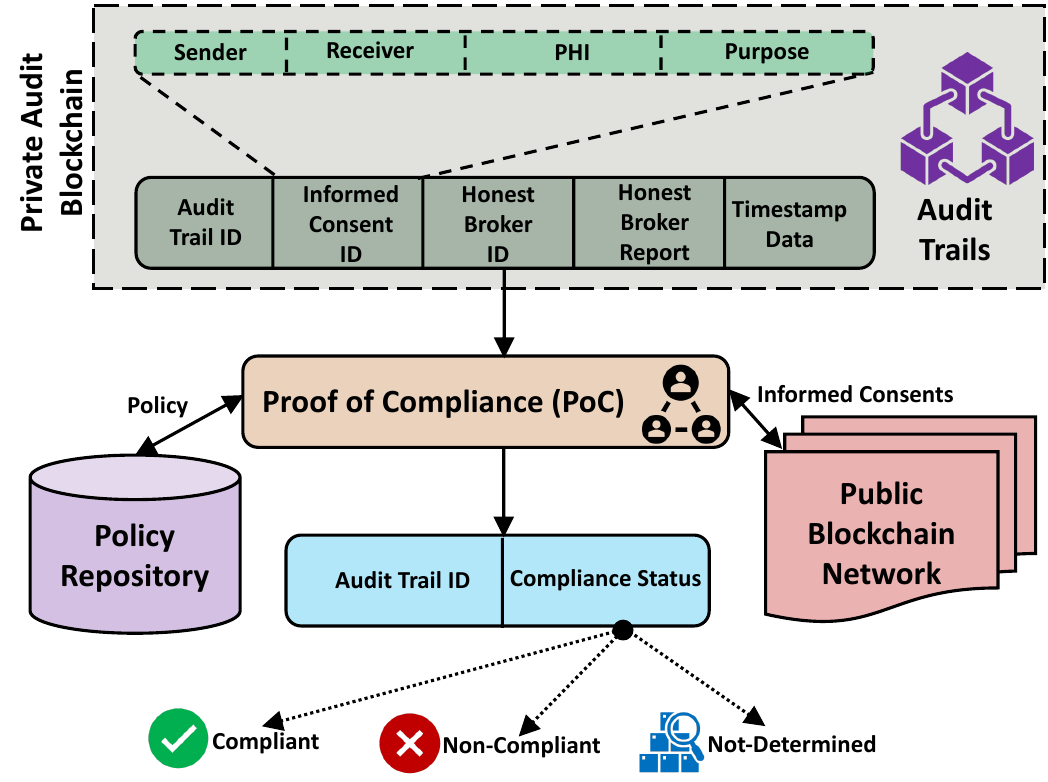}
    \caption{Proof of Compliance (PoC) Transaction Structure.} \label{fig:poc-transaction-structure}
\end{figure}

Figure \ref{fig:poc-transaction-structure} depicts the transaction structure of the Proof of Compliance consensus mechanism. The PoC takes input from an audit trail that contains (i) audit trail ID, (ii) informed consent ID or SIC ID, (iii) honest broker ID, (iv) honest broker report, and (v) timestamp data. Applicable policy and sharing informed consent are retrieved from the policy repository and public blockchain to check the status of each audit trail. After verifying, each auditor node determines the compliance status for each transaction. There are three compliance statuses: (i) \textit{compliant}, which indicates there are no security and privacy policy violations; (ii) \textit{non-compliant} means there is a policy violation, and (iii) \textit{non-determined} defines that required information is not available to check status.

The auditor nodes can be hospitals, various governments, regulatory agencies, insurance companies, business associates, and others. 
They do not store audit trail data and are responsible for maintaining compliance status for each transaction. Reports from all auditor nodes are collected and combined for the final decision. Algorithm \ref{alg:poc-mechanism} shows the core functionalities of PoC: signature verification and order, transaction validation, policy compliance verification, and ledger modification. Due to page constraints, we do not include detailed protocols, communication mechanisms, and synchronization techniques. They are our future research communications with performance evaluations for compliance accuracy measurements, data security and privacy, and others.

\RestyleAlgo{ruled}
\SetKwComment{Comment}{/* }{ */}

\begin{algorithm}[tb]
\scriptsize
    \SetKwInOut{KwData}{Input}
    \SetKwInOut{KwResult}{Output}
    \SetKw{KwBy}{by}
\DontPrintSemicolon

\caption{Proof of Compliance (PoC) Consensus Method.}\label{alg:poc-mechanism}

\KwData{ (i) list of transactions $(Txns)$ and (ii) set of policy $Plcy$}
\KwResult{ (i) list of accepted/rejected transactions $(Txns)$ and
        (ii) list of transactions that are policy compliance}
    
    \textbf{\textit{Initialization}} \vfill 
            (i) $\mathbb{N}_{Order}$: order nodes,
            (ii) $\mathbb{N}_{Validator}$: validator/endorser nodes,
            (iii) $\mathbb{N}_{Audit}$: audit nodes, and
            (iv) $\mathbb{N}_{Committer}$: committer nodes

\textbf{\textit{Signature Verification and Order}} \vfill
         
        $Txn_{Valid} = []$  \textcolor{blue}{\Comment*[r]{accepted transaction list }}
        $Txn_{Invalid} = []$  \textcolor{blue}{\Comment*[r]{rejected transaction list }}

                        \For{$i \gets Txns_{Start}$ \KwTo $Txns_{End}$ \KwBy $1$}{
                                \eIf{$\zeta (PK_i,Tnx_i) == Signed_{Tnx_i}$}{
                                        $Txn_{Valid} \gets Txn_{Valid} + Txn_i$
                                }{
                                        $Txn_{Invalid} \gets Txn_{Invalid} + Txn_i$
                                }
                        }

 \textbf{\textit{Transaction Validation}} \vfill
        $Txn_{Accepted} = []$  \textcolor{blue}{\Comment*[r]{accepted transaction list }}
        $Txn_{Rejected} = []$  \textcolor{blue}{\Comment*[r]{rejected transaction list }}

                        \For{$i \gets {Txn_{Valid}}_{Start}$ \KwTo ${Txn_{Valid}}_{End}$ \KwBy $1$}{
                                \eIf{$\zeta (PK_i,Tnx_i) == Signed_{Tnx_i}$}{
                                        $Txn_{Accepted} \gets Txn_{Accepted} + {Txn_{Valid}}_i$
                                }{
                                        $Txn_{Rejected} \gets Txn_{Rejected} + {Txn_{Valid}}_i$
                                }
                        }

 \textbf{\textit{Policy Compliance Verification}} \vfill

        $Txn_{Compliance} = []$  \textcolor{blue}{\Comment*[r]{compliance transactions }}
        $Txn_{NonCompliance} = []$  \textcolor{blue}{\Comment*[r]{noncompliance transactions}}

        \For{$i \gets {Txn_{Accepted}}_{Start}$ \KwTo ${Txn_{Accepted}}_{End}$ \KwBy $1$}{
                                \eIf{$\zeta (PK_i,Tnx_i) == Signed_{Tnx_i}$}{
                                        $Txn_{Compliance} \gets Txn_{Compliance} + {Txn_{Accepted}}_i$
                                }{
                                        $Txn_{NonCompliance} \gets Txn_{NonCompliance} + {Txn_{Accepted}}_i$
                                }
                        }

 \textbf{\textit{Ledger Modification}} \vfill

        $Txn_{Compliance} = []$  \textcolor{blue}{\Comment*[r]{compliance transactions }}
        $Txn_{NonCompliance} = []$  \textcolor{blue}{\Comment*[r]{noncompliance transactions}}

        \For{$i \gets {Txn_{Accepted}}_{Start}$ \KwTo ${Txn_{Accepted}}_{End}$ \KwBy $1$}{
                                \eIf{$\zeta (PK_i,Tnx_i) == Signed_{Tnx_i}$}{
                                        $Txn_{Compliance} \gets Txn_{Compliance} + {Txn_{Accepted}}_i$
                                }{
                                        $Txn_{NonCompliance} \gets Txn_{NonCompliance} + {Txn_{Accepted}}_i$
                                }
                        }
\end{algorithm}

\section{SIC Provenance Services} \label{sec:consent-services}
Patients need to be provided with the specifics of their given sharing informed consent: who can share what PHI with whom, and for what purposes? Additionally, patients should understand the execution of their consent, including the details of who shares which healthcare data, the timing of these actions, and others. They should also know whether those sharing activities comply with the applicable security and privacy policies, regulatory requirements, industry best practices, contractual obligations, etc. This section outlines the services related to the given and executed consent that patients can access within the proposed framework, provided they have the necessary credentials.  The primary goal of provenance services is to ensure patients receive accurate and comprehensive information and have confidence regarding their given and executed informed consent.

\subsection{Given Consent Services}
\begin{figure}[htb]
    \centering
    \includegraphics[width=\linewidth]{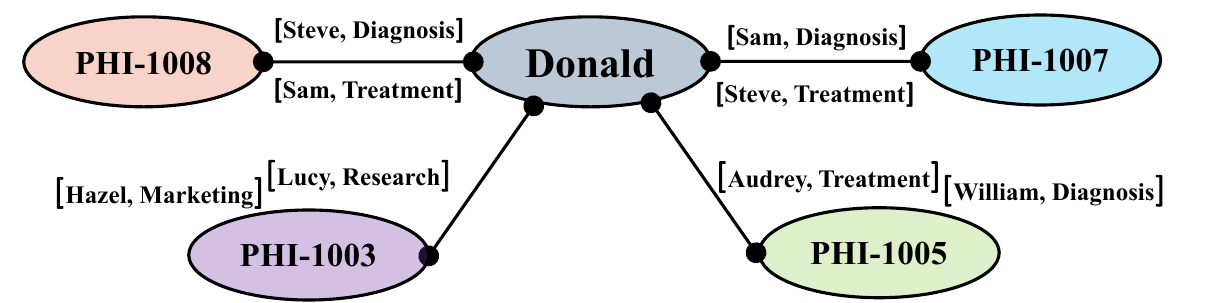}
    \caption{Sender-Oriented Given Consents.} \label{fig:sender-oriented-given-consents}
\end{figure}

\begin{figure}[htb]
    \centering
    \includegraphics[width=\linewidth]{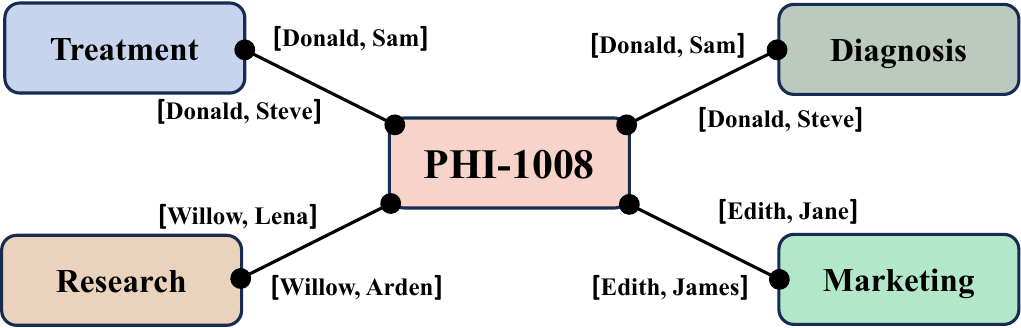}
    \caption{PHI-Oriented Given Consents.} \label{fig:phi-oriented-given-consents}
\end{figure}

In this scope, patients can access the list of all the given consents for sharing healthcare data to date. These consents are in their original state and may or may not be executed for making data-sharing decisions. Patients can see the list where each consent contains information about who the sender is, who the receiver is, what the protected healthcare information is, and the purpose of sharing healthcare data when the sharing informed consent is given. Given consent services can be delivered: \textit{(i) sender-oriented, (ii) receiver-oriented, (iii) PHI-oriented,} and \textit{(iv) purpose-oriented.} For example, patients can have sender-oriented consent services that include all the consents given to a particular sender or a group of senders. Figure \ref{fig:sender-oriented-given-consents} depicts sender-oriented given consents for Donald, who has permission to share PHI with various receivers. Figure \ref{fig:phi-oriented-given-consents} shows the PHI-oriented given consents for health record \textit{PHI-1008}.

\subsection{Executed Consent Services}
\begin{figure}[htb]
    \centering
    \includegraphics[width=\linewidth]{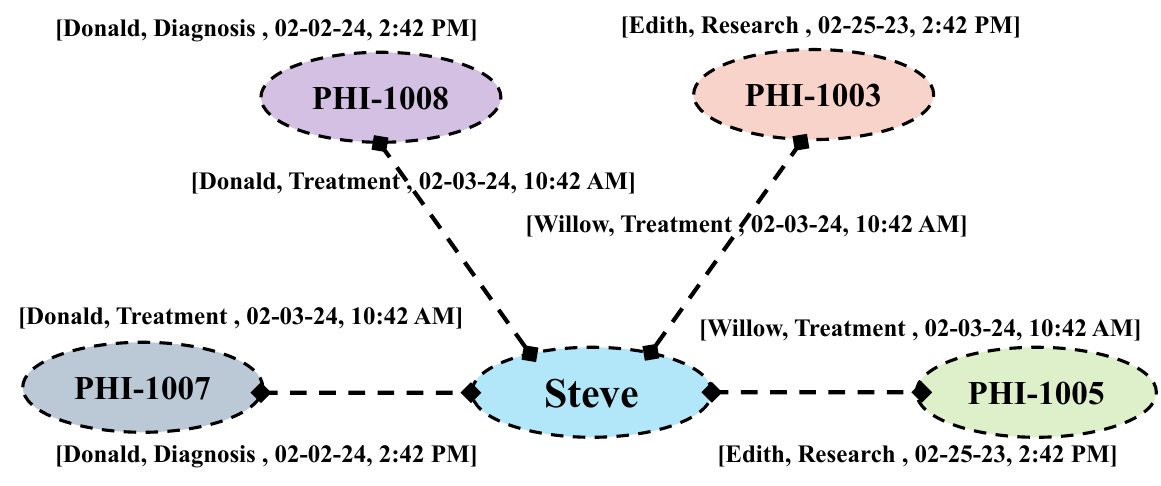}
    \caption{Receiver-Oriented Executed Consents.} \label{fig:receiver-oriented-executed-consents}
\end{figure}

\begin{figure}[htb]
    \centering
    \includegraphics[width=\linewidth]{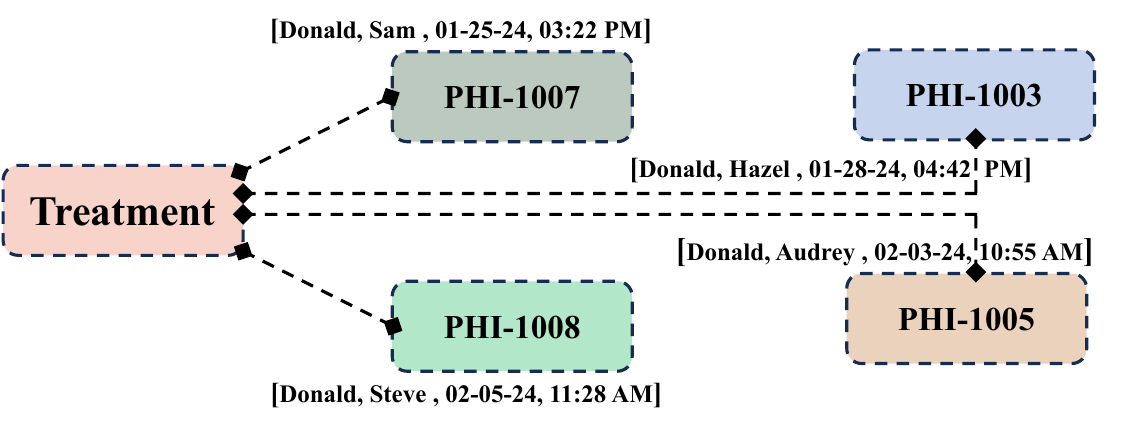}
    \caption{Purpose-Oriented Executed Consents.} \label{fig:purpose-oriented-executed-consents}
\end{figure}

After generation, all consents may or may not be executed to share healthcare data. A consent is executed when a sender wants to share PHI with the receiver when there is a need for it to serve the purpose included in the consent. If consent is executed, other information is stored in addition to the consent, like an honest broker ID, a pertinent policy status that the broker has certified, a timestamp, etc. Executed consent services can be provided: \textit{(i) sender-oriented, (ii) receiver-oriented, (iii) PHI-oriented,} and \textit{(iv) purpose-oriented.} For example, a patient may need to know the executed consent for a particular receiver. Figure \ref{fig:receiver-oriented-executed-consents} shows receiver-oriented executed consents for \textit{Steve} with senders and timestamps. Figure \ref{fig:purpose-oriented-executed-consents} depicts purpose-oriented executed consents for \textit{treatment} with sender, receiver, and timestamp.

\subsection{Service Delivery to Patients}
Patients will interact with the system through interfaces like GUIs or apps supported by wallets like \textit{Coinbase} and \textit{MetaMask} for transaction signing and data access management. These wallets safeguard users' private keys and credentials. The system accommodates various user types, including those requiring tailored interfaces, such as seniors, physically disabled individuals, minors, and others. Healthcare providers may address the specific needs of these diverse users and can develop apps and software to provide services.  Patients' devices and apps are assumed to be secure against unauthorized access, and communication with the blockchain is also protected.

\section{Experimental Evaluation} \label{sec:experimental-proposition}
The Ethereum Virtual Machine (EVM) based three blockchain test networks (\textit{Arbitrum}, \textit{Polygon}, and \textit{Optimism}) are chosen for the experiments. We developed and deployed smart contracts for storing and retrieving PPA integrity and informed consent in test networks. Ethereum's Remote Procedure Call (RPC) API services are employed for deploying smart contracts and performing transactions on these networks \cite{kim2023etherdiffer}. Utilizing public \textit{RPC} eliminates the need to maintain a blockchain node for contract interaction, assuming minimal resource usage (CPU, HDD, bandwidth) on the local machine. We used Metamask wallet to sign and authorize transactions using \textit{ETH} and \textit{MATIC} faucet tokens as gas. Healthcare providers may invest in infrastructure such as blockchain nodes, web interfaces, and mobile applications for seamless service interaction between patients and healthcare systems. Storing informed consent on public blockchains like Ethereum incurs direct monetary costs. Patients, insurance companies, and others can split these costs, like those for doctor visits, medications, and laboratory tests. The following discusses gas consumption and time requirements.

\subsection{Gas Consumption}
\begin{figure}[htb]
    \centering
    \includegraphics[width=\linewidth]{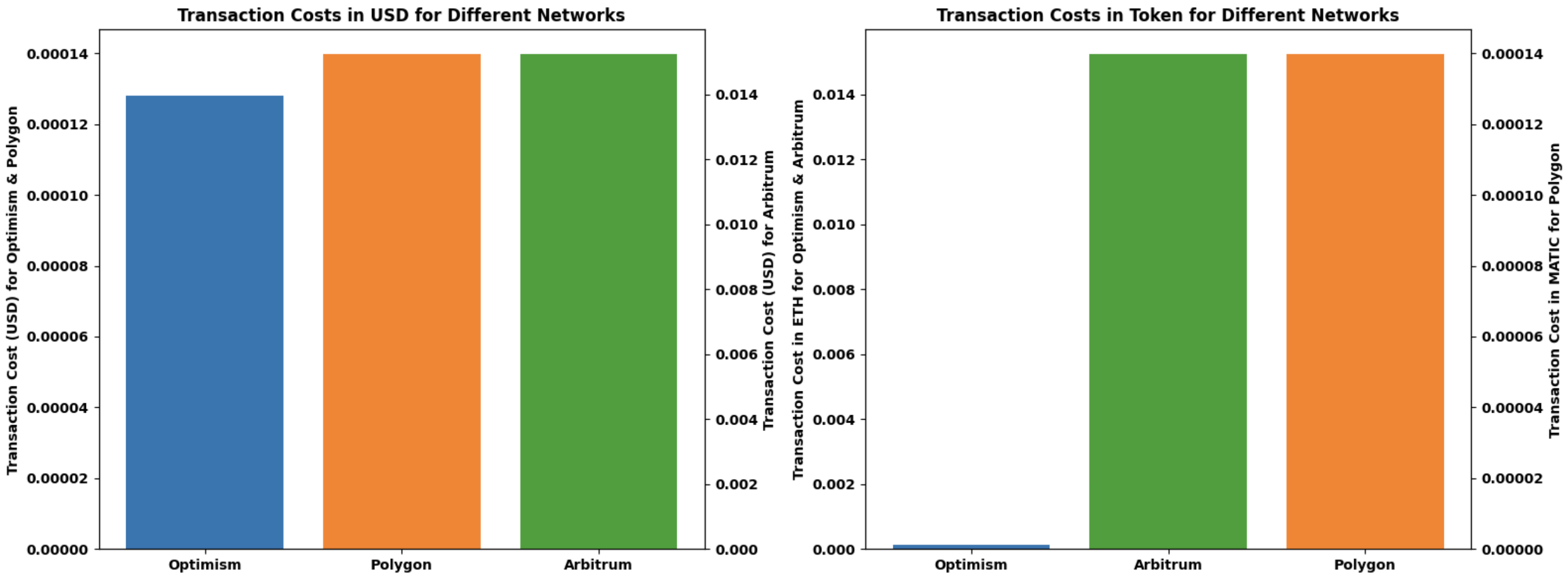}
    \caption{PPA Integrity Storage Cost.} \label{fig:ppa-integrity-cost}
\end{figure}

\begin{figure}[htb]
    \centering
    \includegraphics[width=\linewidth]{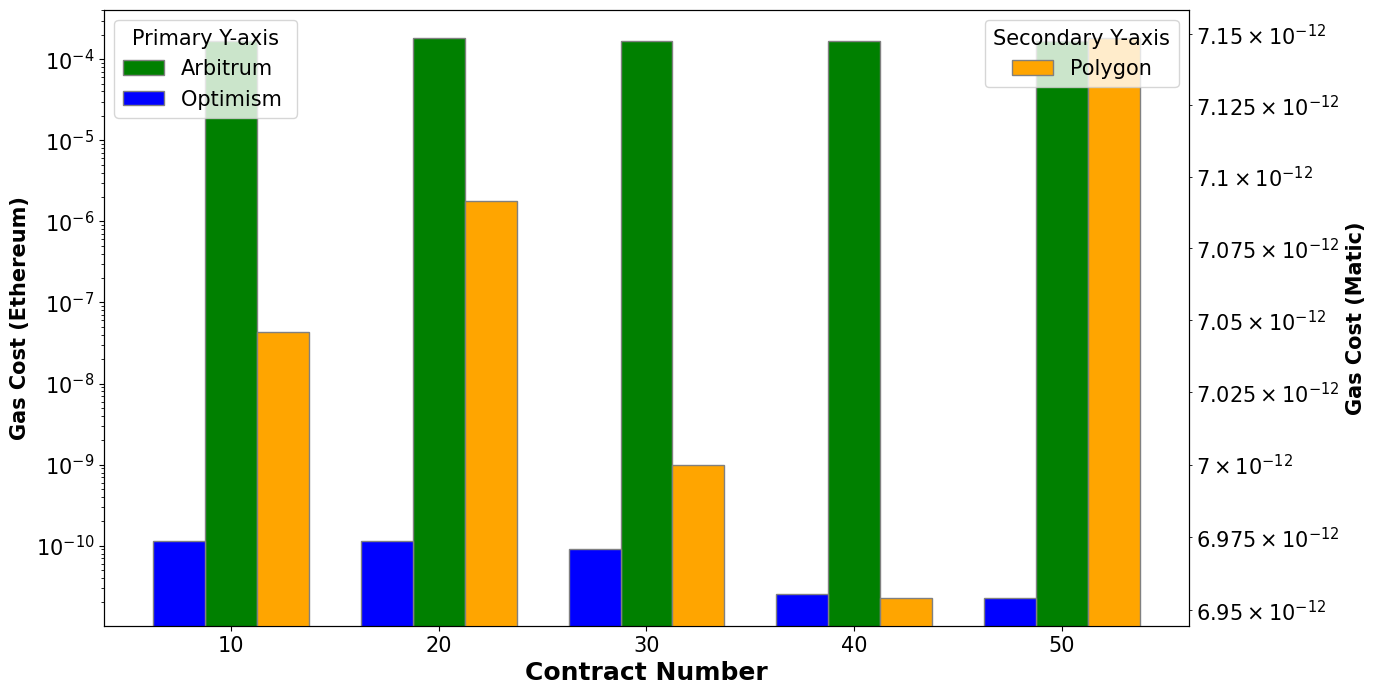}
    \caption{Contract Deployment Gas Cost.} \label{fig:contract-deployment-gas-cost}
\end{figure}

\begin{figure}[htb]
    \centering
    \includegraphics[width=\linewidth]{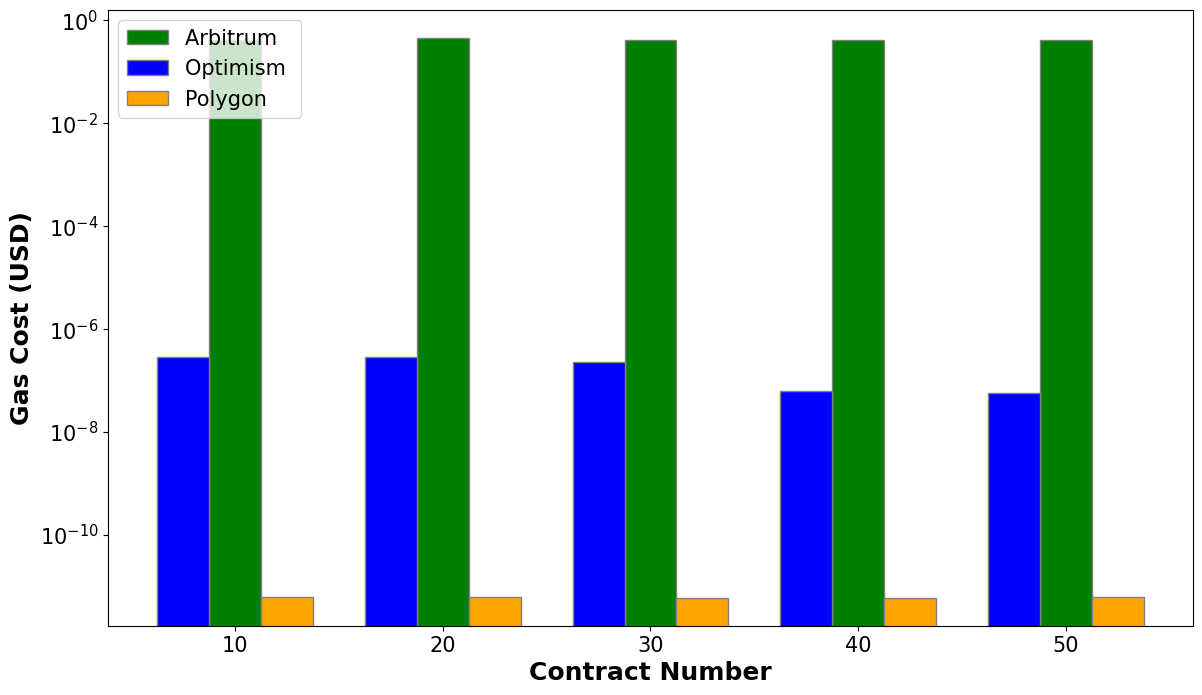}
    \caption{Contract Deployment USD Cost.} \label{fig:contract-deployment-usd-cost}
\end{figure}

\begin{figure}[htb]
    \centering
    \includegraphics[width=\linewidth]{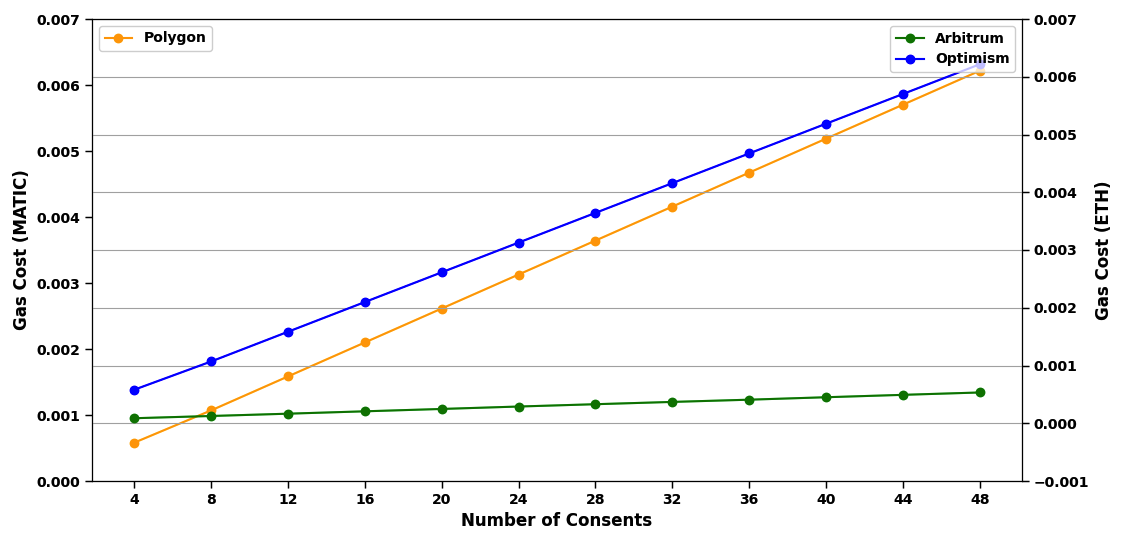}
    \caption{Consent Storage Gas Cost.} \label{fig:all-consent-gas-cost}
\end{figure}

\begin{figure}[htb]
    \centering
    \includegraphics[width=\linewidth]{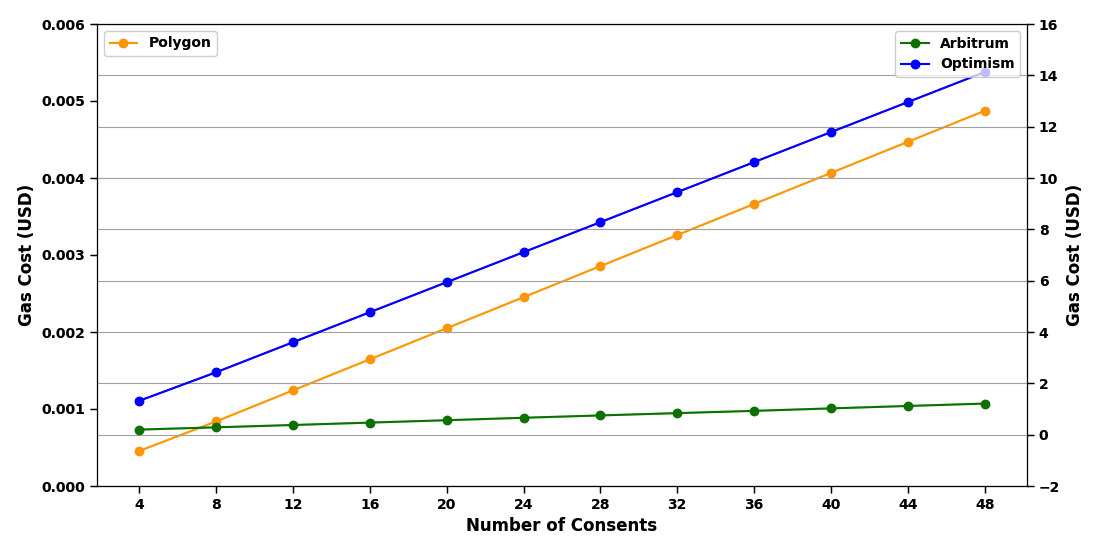}
    \caption{Consent Storage USD Cost.} \label{fig:all-consent-usd-cost}
\end{figure}

Gas is needed for any activity on the Ethereum network involving writing data or changing the state of the blockchain. Smart contract deployment and function calling costs to write data on the blockchain network are considered in this work. A contract is deployed for each patient separately to manage consent-related queries efficiently. The cost of smart contract deployment is proportional to the size of the code \cite{albert2020gasol}. This is a one-time cost for a single-contract deployment. How much it costs to call a function depends on how many times it is called and how much data needs to be stored or changed on the blockchain network. Figure \ref{fig:ppa-integrity-cost}, \ref{fig:contract-deployment-gas-cost}, \ref{fig:contract-deployment-usd-cost}, \ref{fig:all-consent-gas-cost}, and \ref{fig:all-consent-usd-cost} show the contract deployment and consent storage costs in gas (token) and USD for three test networks.  

\subsection{Time Requirements}
Blockchain-based applications require block data writing and reading time requirements. Writing time includes smart contract deployment and data addition. Table \ref{table:consent-writing-time} shows the writing time for various consent numbers for the test networks. The reading time indicates the required time to get data from the block of the blockchain ledger. All the read calls of smart contracts are gas-free. Table \ref{table:consent-reading-time} shows the test network's reading time for various consent numbers. The same smart contracts and consents are used for all test networks. Maintaining a node locally can reduce the reading time from the network where block data can be accessed in real-time. The system continuously synchronizes with the blockchain network to update the ledger data. The providers can maintain local nodes for faster authorizations.

\begin{table}[htb]
\centering
\rowcolors{2}{purple!25}{gray!20}
\caption{Consent Writing Time to Blockchain Network.} \label{table:consent-writing-time}
\scriptsize
\begin{tabular}{|p{0.18\linewidth}|p{0.18\linewidth} |p{0.18\linewidth}| p{0.18\linewidth}|}
\hline
  \rowcolor{purple!50}  \textbf{Consents \#} &  \textbf{Polygon} & \textbf{Arbitrum} & \textbf{Optimism}  \\
\hline
4 & 6.719 Sec & 6.854 Sec & 8.459 Sec \\
\hline
8 & 5.961 Sec & 6.068 Sec & 7.785 Sec \\
\hline
12 & 5.972 Sec & 6.338 Sec & 7.738 Sec \\
\hline
16 & 6.309 Sec & 6.063 Sec & 7.762 Sec \\
\hline
20 & 6.085 Sec & 6.081 Sec & 8.163 Sec \\
\hline
24 & 6.015 Sec & 2.476 Sec & 7.482 Sec \\
\hline
28 & 10.117 Sec & 6.521 Sec & 7.718 Sec \\
\hline
32 & 10.041 Sec & 2.451 Sec & 8.268 Sec \\
\hline
36 & 10.045 Sec & 6.662 Sec & 7.736 Sec \\
\hline
40 & 14.039 Sec & 2.458 Sec & 7.797 Sec \\
\hline
44 & 10.048 Sec & 6.201 Sec & 7.881 Sec \\
\hline
48 & 10.138 Sec & 6.174 Sec & 8.971 Sec \\
\hline
\end{tabular}
\end{table}

\begin{table}[htb]
\centering
\rowcolors{2}{teal!25}{gray!20}
\caption{Consent Reading Time from Blockchain Network.} 
\label{table:consent-reading-time}
\scriptsize
\begin{tabular}{|p{0.18\linewidth}|p{0.18\linewidth}|p{0.18\linewidth}|p{0.18\linewidth}|}
\hline
\rowcolor{teal!60} \textbf{Consents \#} & \textbf{Polygon} & \textbf{Arbitrum} & \textbf{Optimism} \\
\hline
4 & 0.426 Sec & 0.234 Sec & 0.399 Sec \\
\hline
8 & 0.366 Sec & 0.201 Sec & 0.423 Sec \\
\hline
12 & 0.337 Sec & 0.239 Sec & 0.425 Sec \\
\hline
16 & 0.346 Sec & 0.259 Sec & 0.423 Sec \\
\hline
20 & 0.327 Sec & 0.288 Sec & 0.442 Sec \\
\hline
24 & 0.344 Sec & 0.241 Sec & 0.579 Sec \\
\hline
28 & 0.358 Sec & 0.221 Sec & 0.536 Sec \\
\hline
32 & 0.361 Sec & 0.288 Sec & 0.495 Sec \\
\hline
36 & 0.401 Sec & 0.225 Sec & 0.512 Sec \\
\hline
40 & 0.36 Sec & 0.206 Sec & 0.482 Sec \\
\hline
44 & 0.361 Sec & 0.233 Sec & 0.462 Sec \\
\hline
48 & 0.522 Sec & 0.224 Sec & 0.434 Sec \\
\hline
\end{tabular}
\end{table}

\section{Conclusions} \label{sec:conclusion-future-directions}
Sharing patient health data is beneficial for improving medical care, diagnosis, and other essential services. However, keeping this information private and secure is important. Different policies from various authorities help ensure the privacy and security of this health data. Complying with these policies ensures that safety measures are working. Getting patients' informed consent is also critical to protecting their privacy and giving them control over sharing their information. Patients need to understand fully how their data is shared.  Patients should also feel confident that strong safeguards are in place to protect their data. Using smart contracts to manage patient consent is a promising way to securely and privately share health data. These systems let patients control their health records and agree to how doctors and others use them. Blockchain technology improves these systems by providing security, efficiency, decentralization, transparency, and immutability. This enhances the trustworthiness and responsibility of sharing healthcare data among everyone involved. 

Looking forward, our objective is to provide functional mechanisms for essential consent management operations for data sharing and enhancing patient care and services. Management operations generate, modify, withdraw, expire, and archive consent. Improper consent can cause sensitive data disclosure or prevent getting services. Consent generation must be done carefully. It is necessary to modify a given consent due to improper components like the receivers or purposes. In this situation, a modified new consent must be deployed, while the old consent must be moved to the achieving repository.

\section*{Acknowledgements}
This work was partially supported by the U.S. National Science Foundation under Grant No. 1822118 and 2226232, the member partners of the NSF IUCRC Center for Cyber Security Analytics and Automation – Statnett, AMI, NewPush, Cyber Risk Research, NIST, and ARL – the State of Colorado (grant \#SB 18-086), and the authors’ institutions. Any opinions, findings, conclusions, or recommendations expressed in this material are those of the authors and do not necessarily reflect the views of the National Science Foundation or other organizations and agencies.

\bibliographystyle{IEEEtran}
\bibliography{main}
\end{document}